\documentclass[11pt]{article}

\usepackage[pdftex]{graphicx}
\usepackage{amsmath,amssymb}
\def\lromn#1{\uppercase\expandafter{\romannumeral#1}}

\begin{document}

\vspace*{2cm}
\begin{center}
\begin{Large}
\textbf{
Precision electroweak shift of  muonium hyperfine splitting
}

\end{Large}

\vspace{2cm}

\begin{large}
T. Asaka,
M. Tanaka, 
K. Tsumura,
and
M.~Yoshimura

\vspace{0.5cm}
Department of Physics, Niigata University, 950-2181 Niigata, Japan \\
Department of Physics, 
         Osaka University, Toyonaka, Osaka 560-0043, Japan \\
Department of Physics, Kyoto University, 606-8502 Kyoto, Japan \\
Research Institute for Interdisciplinary Science,
 Okayama University 
700-8530 Japan 
\end{large}
\end{center}

\vspace{6cm}
\begin{center}
\begin{Large}
{\bf ABSTRACT}
\end{Large}
\end{center}

Electroweak second order shifts of muonium ($\mu^+e^-$ bound state) energy levels are
calculated for the first time.
Calculation starts from on-shell one-loop elastic $\mu^+ e^-$ scattering amplitudes in the center of mass frame,
proceed to renormalization and to derivation of muonium matrix elements
by using the momentum space wave functions.
This is a reliable method unlike the unjustified four-Fermi approximation in the literature.
Corrections of order $\alpha G_F$ (with $\alpha \sim 1/137$ the fine structure
constant and $G_F$ the Fermi constant) and of order
$\alpha G_F /(m_Z a_B)$ (with $m_Z$ the Z boson mass and $a_B$ the Bohr radius) are derived from
three classes of Feynman diagrams, Z self-energy, vertex and box diagrams.
The ground state muonium hyperfine splitting is given in terms of the only experimentally unknown parameter,
the smallest neutrino mass.
It is however found that the neutrino mass dependence is very weak, making
its detection difficult.

\newpage
 {\bf Introduction} \hspace{0.3cm}
Electroweak correction to atomic energy levels has been found an important tool
to provide parity violation of atomic force, which was proved by
atomic parity violation experiments \cite{wieman}, giving the weak mixing
angle consistent with the one determined by neutral current weak interaction phenomena in the high
energy frontier. 
This correction is of the first order of the Fermi coupling constant 
$G_F \sim 1.166 \times 10^{-5}$GeV$^{-2}$.
Second order electroweak effects have often been stated
to be negligible, presumably due to a misconception that
it might be of order $G_F^2$.
But actually in the renormalizable electroweak theory second order correction
is of order $\alpha G_F$  for flavor diagonal parts.
Flavor changing effects of order $G_F^2 \delta m^2$ with $\delta m^2$
mass differences of quarks or leptons
\cite{fcn in ew} do not give
a good hint because this involves off-diagonal matrix elements
unlike our interest in flavor-diagonal parts.
The question then arises how large the coefficient of $\alpha G_F$  order is.
The answer to this question is given only after a full body
of one-loop calculation, which we shall address to.

A different motivation for second order electroweak calculation is how the long range force mediated
by nearly massless neutrino-pairs emerges in atomic energy shifts.
The long range force caused by the neutrino pair exchange between
charged fermions has been studied in the literature \cite{pair-force 1},
\cite{pair-force 2}, \cite{pair-force 3} since the old days of four-Fermi theory of weak interaction.
We shall demonstrate that our full-body electroweak calculation
does not justify the four-Fermi approximation,
and estimate \cite{pair-force 4}  of atomic energy shifts based on this
approximation is dubious.

We concentrate on purely leptonic bound systems, since spectroscopy of
hydrogen and munonic hydrogen both suffers from the
proton structure effect, recently of much debate.
But if strong interaction is well under  control, applications to these systems
should be possible.

We use the natural unit of $\hbar = c = 1$ unless otherwise stated.

\vspace{0.5cm}
{\bf First order electroweak effect and basic framework for one-loop calculation} \hspace{0.3cm}
$SU(2) \times U(1)$ electroweak gauge theory \cite{electroweak} is renormalizable, which
means that their higher loop effects to muonium energy levels are calculable from
the first principles.
A non-trivial part of calculation from higher order loop diagrams 
is renormalization, whose method is however well established \cite{ew theory}.
We work in the Feynman gauge within a more general framework 
of $R_{\xi}$ gauge \cite{r_xi gauge}.
This considerably  simplifies the burden of  calculation.

We introduce finite neutrino masses in the electroweak theory which is however
generated by physics beyond the standard electroweak theory.
There are two kinds of neutrino masses, the Dirac and the Majorana types.
Since distinction of these two cases is found difficult in muonium (Mu) spectroscopy, 
we shall present calculations in the easier case of Dirac neutrino.
We  regard the vanishing smallest neutrino mass  is
highly unlikely, and assume that all neutrinos are massive
regardless of their small  values.

There exists the first order electroweak shift of order $G_F$.
Contribution of Z boson exchange to HFS arises from the fact
that Z boson has a coupling of the axial vector current along with the vector current.
Unlike the particle velocity operator for the vector current
the spatial part of axial vector current is the spin current.
Z boson exchange thus gives rise to a spin-spin interaction $\propto  \vec{S}_{\mu}\cdot  \vec{S}_e$
between $\mu^+$ and $e^-$, which is the same operator form as the one
responsible for ordinary magnetic HFS.
The predicted HFS of 1s Mu is $- \sqrt{2} \alpha^3 G_Fm_e^3 (1 + m_e/m_{\mu})^{-3}/\pi =- $  65 Hz,
 which sets a scale of weak effects.

\begin{figure*}[htbp]
 \begin{center}
\includegraphics{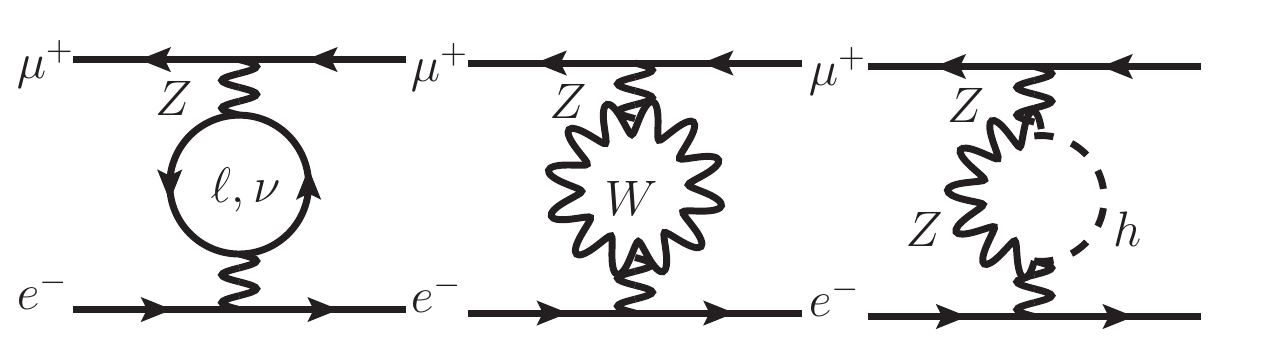}
   \caption{
Z boson self-energy diagrams of neutrino $\nu$, charged lepton $l$  and $W H$ pairs.
}
   \label {muonium HFS 1}
 \end{center} 
\end{figure*}

\begin{figure*}[htbp]
 \begin{center}
\includegraphics{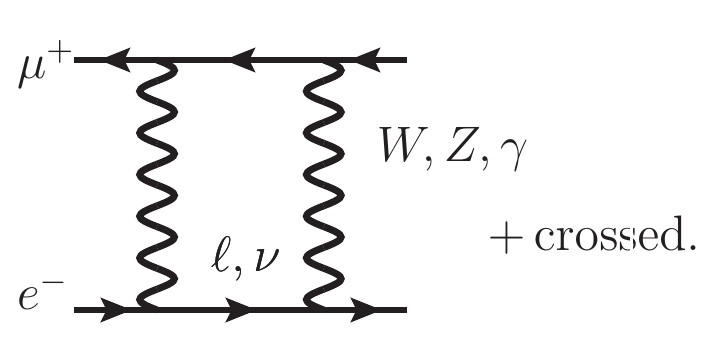} 
\vspace{1.5cm}
   \caption{
t-channel exchange consisting of pairs $(W^+ W^-)\,, (Z, Z)\,, (Z, \gamma)$
}
   \label {muonium HFS 3}
 \end{center} 
\end{figure*}

There are three classes of Feynman diagrams at one loop level as shown in Fig(\ref{muonium HFS 1})
$\sim$  Fig(\ref{muonium HFS 3}) (and similar ones) that
contribute to atomic force between two charged fermions like $\mu^+ e^-$:
they are called Z-boson
self-energy, vertex diagrams and box diagrams.
Our calculation starts from low energy elastic scattering amplitudes in the center
of mass frame,  proceed to renormalization for self-energy and vertex, derive
the force potential by Fourier-transforming  the momentum 
transfer $\vec{q}$ to the position vector $\vec{r}$,
and finally calculate relevant atomic matrix elements, using relevant wave functions.
The procedure is standard and there seems no other rigorous way to
extract atomic energy shifts.
Convergence of integrals is found better if one uses wave functions in
the momentum space, hence we bypass derivation of the force potential
which is much more complicated than those shown in \cite{pair-force 1} 
$\sim$ \cite{pair-force 3}, valid only in the four-Fermi approximation.
In the four-Fermi approximation all heavy internal boson ($Z\,, W$) lines are
shrunk to a point, and there is no structure difference in all of one-loop Feynman diagrams.
We demonstrate explicitly below that this approximation is wrong
and three classes of Feynman diagrams give different dependences on
relevant parameters such as $m_Z, m_W, \alpha$ and neutrino masses.

\begin{figure*}[htbp]
 \begin{center}
\includegraphics{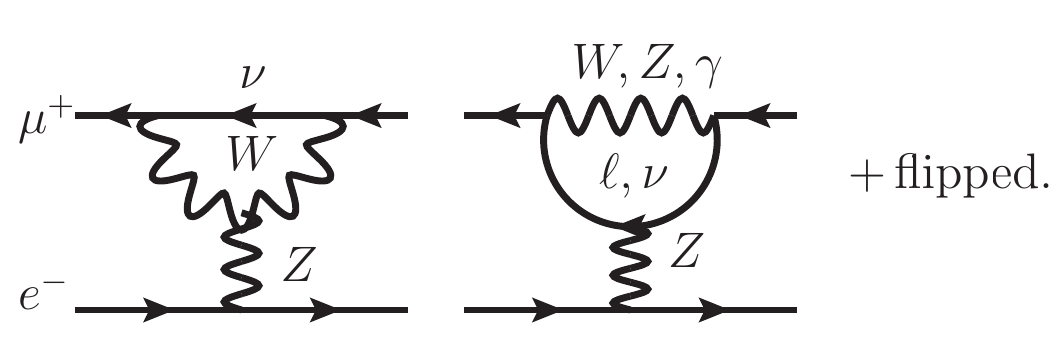}
   \caption{
Vertex diagrams with $Z$ attached to $e^-$ line containing $W^+\!$-$\!W^-\!$-$\!Z$ coupling in the left. 
The triangle vertex may also be attached to electron line.
}
   \label {muonium HFS 2}
 \end{center} 
\end{figure*}

The ghost contribution is suppressed by a mass factor when it is coupled to
fermions.

\vspace{0.5cm}
{\bf Vacuum polarization and self-energy diagram of Z boson} \hspace{0.3cm}
The force potential may be derived by Fourier-transforming the elastic scattering amplitude
${\cal A}_{EW} (\vec{q}) = \Pi^R (\vec{q})/(\vec{q}\,^2 + m_Z^2)^2$.
Here $\Pi^R$ is space-space component of vacuum polarization tensor.
The space-space component gives in the Feynman gauge of Z propagator
the required structure  $\propto  \vec{S}_{\mu}\cdot  \vec{S}_e$ of HFS. 
It is found that the Fourier transform is a complicated function of position.
We shall bypass this derivation and calculate diagonal matrix elements using
the momentum-space wave function of muonium.
According to \cite{bs}, the Schroedinger equation including the electroweak correction
may be written down as an integral equation in the momentum space,
 and one derives the energy shift caused by the electroweak term ${\cal A}_{EW} (\vec{q})$:
\begin{eqnarray}
&&
\Delta_{1s HFS}^{i p} =
- \frac{ (g^2 + (g')^2)^2} {32 } J_i
\,, \hspace{0.5cm}
J_i = \frac{1}{ (2\pi)^3} \int d^3 p d^3q \, \psi_c^*(\vec{p} + \vec{q})  
{\cal A}_{EW}^i(\vec{q} ) \psi_c (\vec{p})
\,, 
\\ &&
\hspace*{-0.5cm}
{\cal A}_{EW}^i(\vec{q}) = c_i \frac{\Pi^R_i(q^2)}{ (q^2 + m_Z^2)^2}
\,, \hspace{0.5cm}
c_{\nu} = 1
\,, \hspace{0.5cm}
c_l = 1 - 4 \sin^2 \theta_w + 8 \sin^4 \theta_w 
\sim 0.50 c_{\nu}
\,,
\end{eqnarray}
where $\psi_c (\vec{p})$ is the momentum-space wave function under the Coulomb potential.
$g\,, g'$ are two gauge coupling constants and $\theta_w = {\rm arctan}\, g'/g$
is the weak mixing angle with $\sin^2 \theta_w \sim 0.239$ at low energies.

We renormalize  the transverse polarization tensor at a space-like momentum transfer 
$q^2= q_0^2 - \vec{q}\,^2 \equiv -\mu^2 < 0$ to derive $\Pi^R_i(q^2)$,
and impose the mass shell condition at $q^2= m_Z^2$.
The renormalization point $\mu^2$ should be chosen to match physics in question,
and for our atomic calculation the appropriate point is $\mu^2 = O(\alpha^2 m_e^2)$,
the inverse of atomic size squared.
For the neutrino-pair exchange and the lepton-pair and quark-pair exchange
this results in, to a good approximation,
\begin{eqnarray}
&&
\Pi^R_{\nu}(\vec{q}\,^2) =  - \frac{2}{3 (4\pi)^2} q^2 \ln (q a_{\mu})^2
\,, \hspace{0.5cm}
\Pi^R_{l, q}(\vec{q}\,^2) =   \frac{2}{15 (4\pi)^2} \frac{q^4}{m_{l,q} ^2}
\,.
\end{eqnarray}
The massless neutrino limit is well defined.
The ground state muonium  HFS is calculated as
\begin{eqnarray}
&&
\Delta_{1s HFS}^{i p} = - \frac{\alpha G_F}{\sqrt{2}\pi^2 \sin^2 \theta_w \cos^2 \theta_w} 
a_{\mu}^{-3 } c_i K_i
\sim - 3.8 \,{\rm Hz}\, c_i K_i
\,, \hspace{0.5cm}
 a_{\mu} = \frac{1 + \frac{m_e}{m_{\mu}} }{\alpha m_e}
\,,
\\ &&
K_{\nu} = \frac{1 }{4 m_Z a_{\nu} } (\ln (m_Z a_{\mu} ) -1 )
\,, \hspace{0.5cm}
K_l = - \frac{0.50}{15m_Z a_{\nu} (m_l a_{\mu})^2} 
\,.
\label {integrand f2}
\end{eqnarray}
The contribution is suppressed by an extra large factor $m_Z a_{\mu}  $ from $O(\alpha G_F)$,
and its value is insensitive to  neutrino mass.
Numerically, we find  $\Delta^{\nu p} = - 2.8 \times 10^{-2}$ mHz adding three neutrino pairs,
and the lepton-pair and the quark-pair contributions are more than six orders of
magnitudes smaller.

Thw weak boson pair contribution is calculated as
\begin{eqnarray}
&&
\Delta_{1s HFS}^{W p} = - \frac{3 \alpha G_F}{80\sqrt{2}\pi^2 m_Z a_{\mu} \cos^2 \theta_w} 
a_{\mu}^{-3 } \sim - 6.2 \times 10^{-6}\, {\rm mHz}
\,.
\end{eqnarray}

\vspace{0.5cm}
 {\bf Vertex and box diagrams} \hspace{0.3cm}
The relevant integral for vertex operator is, after dimensional regularization
and renormalization, $B^R(q) = B(q) - B(0)$
\begin{eqnarray}
&&
B(q)= i \int \frac{d^4 k}{(2\pi)^4} \frac{{\cal N}} { (k^2 - \Delta(q) \,)^3}
= - \frac{1}{(4\pi)^2} \int_0^1 dx \int_0^{1-x} dy  \left( \ln \Delta(q) + \frac{m_i^2}{ \Delta(q) } y^2 \right)
\,, 
\\ &&
\Delta = M^2 y + \vec{q}\,^2 x (1-x - y) - m_l^2 y (1- y) + m_i^2 (1-y)
\,.
\end{eqnarray}
To leading $M^2 =  m_W^2\,, m_Z^2$ order,
\begin{eqnarray}
&&
B^R(q) \simeq -\frac{q^2 }{M^2 (4\pi)^2} (\frac{1}{ 3}\ln \frac{M }{q } + \frac{5}{ 36}) 
\,.
\end{eqnarray}
The ground state Mu HFS is calculated:
\begin{eqnarray}
&&
\Delta_{1s\, HFS}^{W\nu} = 
\frac{\alpha G_F  }{48 \sqrt{2} \pi^2 \sin^2 \theta_w \cos^2\theta_w } \frac{a_{\mu}^{-3} }{m_Z a_{\mu} }
( \frac{5 }{12 } - \ln \frac{m_Z }{m_W } )
\,,
\\ &&
\Delta_{1s\, HFS}^{Z l} = 
- \frac{\alpha G_F  }{96 \sqrt{2} \pi^2 \sin^2 \theta_w \cos^2\theta_w } \frac{a_{\mu}^{-3} }{m_Z a_{\mu} }
\frac{5 }{12 } 
\,.
\end{eqnarray}
These vertex contributions are doubled adding the vertex attached to another side of lepton.
Numerically, it is of order, $1.4  \times 10^{-5}$ mHz, for $W-\nu$ vertex,
and $- 1.0 \times 10^{-5}$mHz for $Zl $ vertex..


The vertex contribution of triangle $Z-\gamma-\gamma$ has infrared divergence problem,
which shall be discussed elsewhere along with other diagrams containing photon.

The box diagram is simplest to calculate, since renormalization is not
required, resulting in
\begin{eqnarray}
&&
\Delta_{HFS, b}^{(1s)} =-  \frac{G_F \alpha a_{\mu}^{-3} }{2 \sqrt{2} \pi^2 } \frac{1}{ \sin^2 \theta_w} \left( 1 
+ \frac{20}{  \cos^2\theta_w }(  \sin^2 \theta_w - \frac{1}{4} )^2   \right)
\sim-  160 \,{\rm mHz} 
\,.
\end{eqnarray}
This contribution of order $\alpha G_F$ is the largest among three classes of Feynman diagrams.

\vspace{0.5cm}
{\bf Summary for the ground state Mu HFS}\hspace{0.3cm}
Our result of 1s Mu HFS is ($- 160 + (- 2.8 + 0.4) \times 10^{-2}  $)mHZ.
Three numbers refer to major contributions from box, self-energy and vertex diagrams.
Neutrino mass dependence appears with a factor $m_{\nu}^2 \ln m_{\nu}^2$ besides $\alpha G_F$,
hence this is a very small correction to HFS.


The present status of 1s Mu HFS is as follows:
latest and best experimentally measured 1s Mu HFS is 4 463 302 776(51) Hz,
\cite{muonium hfs ex}, while theoretical prediction is 4 463 302 891(272) Hz
\cite{muonium hfs th}, \cite{nomura}.
A goal of QED higher order calculations is around 10 Hz \cite{qed}.
According to \cite{shimomura}, the level of HFS accuracy of order 10 Hz
is feasible in a forthcoming experiment.
These inputs are sufficient to discover the major electroweak effect $-$ 65 Hz.

Since our work indicates that the second order electroweak shift to 1s Mu HFS
is around $400 $ smaller than the first order shift,
the  next goal beyond $-65$ Hz is to look for new physics.
Any discrepancy may be attributed to new physics beyond the standard electroweak theory
since second order effects are negligible,
provided QED higher order corrections are better understood.
Assuming a comparable coupling with electroweak theory,
new physics  may be searched to energy scale of a few TeV.

\vspace{0.5cm}
 {\bf Acknowledgements}

We thank K. Shimomura for explaining the status of on-going muonium HFS
experiment at KEK.
This research was partially
 supported by Grant-in-Aid 
17K05410(TA), 18H05543(KT), 17H05405(MT), 17H02895(MY) from the
 Ministry of Education, Culture, Sports, Science, and Technology.


\begin{thebibliography}{99}
\bibitem{fcn in ew}
Suppression of flavor changing neutral current was instrumental to the
success of electroweak theory.
This was realized by Glashow-Illiopoulos-Minami (GIM) mechanism
that led to the unitary matrix of Kobayashi-Maskawa model
of quark mixing and small lepton flavor violating rates.
Diagonal parts whose example is $g-2$
of muon are however free of this suppression
and there is no reason against order $\alpha G_F$ corrections.
\end{thebibliography}
\end{document}